\documentclass[review]{elsarticle}

\usepackage{lineno,hyperref}
\usepackage{graphicx}                       
  
\usepackage{color}                          %
\usepackage{amsmath}                        %
\usepackage{amssymb}                        %
\usepackage[below]{placeins}                %
\usepackage{flafter}                        %
\usepackage{multirow}                       %
\usepackage{booktabs}                       %
\usepackage{longtable}                      %
\usepackage{tabularx}                       %
\usepackage{subfigure}                      %
\usepackage[amsmath,thmmarks,hyperref]{ntheorem}  %

\usepackage[ruled,lined,linesnumbered]{algorithm2e}       %
\usepackage{algorithmic}
\usepackage{makecell}

\modulolinenumbers[5]

\begin{document}

\begin{frontmatter}

\title{Analysis of 5G academic Network based on graph representation learning method }
\tnotetext[mytitlenote]{Fully documented templates are available in the elsarticle package on 96
 \href{http://www.ctan.org/tex-archive/macros/latex/contrib/elsarticle}{CTAN}.}

\author[mymainaddress,mysecondaryaddress]{Xiaoming Li}
\ead{lxm696@tju.edu.cn}

\author[Mythirdaddress,Myfourthaddress]{Guangquan Xu\corref{mycorrespondingauthor}}
\cortext[mycorrespondingauthor]{Corresponding author}
\ead{losin@tju.edu.cn}

\author[mymainaddress,mysecondaryaddress]{Wei Yu\corref{mycorrespondingauthor}}
\ead{weiyu@zyufl.edu.cn}

\author[Myfourthaddress]{Pengfei Jiao}
\ead{pjiao@tju.edu.cn}
\author[Myfifthaddress]{Xiangyu Song}
\ead{xiangyu.song@deakin.edu.au}

\address[mymainaddress]{Zhejiang Yuexiu University of Foreign Languages, School of International Business, Shaoxing, China}
\address[mysecondaryaddress]{Shaoxing Key Laboratory of Intelligent Monitoring and Prevention of Smart City, Shaoxing, China}
\address[Mythirdaddress]{School of Big Data, School of Big Data, Qingdao Huanghai University, Qingdao, China}
\address[Myfourthaddress]{Tianjin Key Laboratory of Advanced Networking(TANK), College of Intelligence and Computing, Tianjin University, Tianjin, China}
\address[Myfifthaddress]{School of IT, Faculty of Science, Engineering and Built Environment, Deakin University, Geelong, VIC 3220, Australia}

\begin{abstract}
 With the rapid development of 5th Generation Mobile Communication Technology (5G),   the diverse forms of collaboration and extensive data in academic social networks constructed by 5G papers make the management and analysis of academic social networks increasingly challenging. Despite the particular success achieved by representation learning in analyzing academic and social networks, most present presentation learning models focus on maintaining the first-order and second-order similarity of nodes. They rarely possess similar structural characteristics of spatial independence in the network. This paper proposes a Low-order Network representation Learning Model (LNLM) based on Non-negative Matrix Factorization (NMF) to solve these problems. The model uses the random walk method to extract low-order features of nodes and map multiple components to a low-dimensional space, effectively maintaining the internal correlation between members. This paper verifies the performance of this model, conducts comparative experiments on four test datasets and four real network datasets through downstream tasks such as multi-label classification, clustering, and link prediction. Comparing eight mainstream network representation learning models shows that the proposed model can significantly improve the detection efficiency and learning methods and effectively extract local and low-order features of the network. 
\end{abstract}

\begin{keyword}
\texttt{5G Academic Network}\sep {Graph Representation Learning} \sep {Non-negative Matrix Factorization}\sep {Random Walk}\sep {multi-label classification}

\end{keyword}

\end{frontmatter}





\section{Introduction}
With the rapid development of 5G technology,  the analysis of 5G academic, social networks is of great academic significance and helps to guide future scientific development. In recent years, researchers have studied 89 million papers published between 1900 and 2015 and found that the collaboration rate between authors has increased 25-fold in the last 116 years. In addition, as measured by the top $1\%$ of cited papers, more than $90\%$ of the world's leading innovations were done by the organization in the early 21st century, which is almost four times as many as in the early 20th century\cite{hoang2017consensus}. This suggests that how to find valuable collaborators has become very important\cite{sun2011co}, and further researchers have demonstrated that scientists or teams with tightly connected collaborative networks tend to generate more research\cite{li2014acrec}.

Research collaboration is a crucial mechanism through which knowledge and capacity, new ideas and research approaches are linked. Specifically,  collaboration connects different talent sets to generate new research. The academic collaboration network can be constructed by taking other authors or institutions as nodes of the network structure and the cooperative relationships between authors or institutions as the links of the network structure. Based on the constructed academic cooperation network, the degree of network structure, node intimacy, intermediation, and PageRank can be analyzed. However, the deeper characteristics of the cooperation network can not be analyzed. In addition, the number of papers or patents published in academic cooperation is large, and the collaboration among authors has a particular tendency, and the partners of authors are relatively few and fixed, resulting in a very sparse link matrix representing the cooperation network of authors. Recent advances in complex network representation learning\cite{DBLP:conf/www/0004ZMK20,2021Latent} allow us to mitigate the challenges of large scale and sparsity in academic collaboration networks. The benefits and effectiveness of complex network representation learning have been demonstrated in many multi-tasks, such as node classification, link prediction, and community detection\cite{DBLP:journals/corr/abs-1806-01261,chen2020efficient,DBLP:journals/tbd/ZhangYZZ20}. In recent years, a number of tasks related to academic collaboration networks have emerged, such as recommending author reviewers or future cooperative research institutions, predicting scholars' research interests\cite{makarov2019dual}, and mining scholars' collaborative model, representing the combination of learning model and scholar attribute of academic collaboration network\cite{DBLP:journals/tcss/YuXL19,2014Academia,2021Measuring}. They argue that student relationships should be represented by co-authors and the research expertise described in the academic paper. Therefore, scholars should be represented in various attributes that reflect their existing collaborative structures and theoretical knowledge. However, these symbolic learning models ignore spatially independent structural similarity characteristics in academic cooperation networks.

Therefore, in this paper, the representation learning method of a complex network is applied to the academic cooperative network. While retaining the structural features of the network as much as possible, the low-dimensional representation of different nodes in the network can be trained to learn both local structural elements in the network and spatially independent structural similarity features in the network. 
Finally, the embedding results are applied to complex network tasks such as classification and clustering to excavate the cooperation law and patterns and trends of cooperation trend between schools and enterprises based on the characteristics of the network underlying structure.

In summary, the contribution of this paper can be described as follows: (1) We propose a low-order network representation learning model, namely  LNLM, to map multiple components into a low-dimensional space while maintaining internal dependencies between components. (2) Comparing to other representation learning model, our proposed model preserves spatially independent structural similarity characteristics in the network. (3) By evaluating the effects on multiple datasets, including the 5G academic social network, the method proposed in this paper is shown to outperform existing higher-order dynamic network methods.

The paper is organized as follows: related work is discussed in Section 2; the proposed method is introduced in Section 3; experimental results are presented in Section 4; Section 5 concludes this paper.

\section{Related work}
\subsection{Academic Collaborative Research}
Academic network analysis effectively deals with explosive educational information and diverse associations in the academic big data environment. As a specific type of social network, academic collaboration networks play an increasingly important role in disseminating academic data and expanding academic collaboration. There are many ways to build an educational network, including co-citation, cooperation, similar research contents and everyday academic activities. In the analysis of authors and collaborators, Wang et al.\cite{DBLP:journals/tcss/WangLYKX19}and Evans et al.\cite{DBLP:journals/scientometrics/EvansLP11}found an apparent homogeneity in academic cooperation, as scientists tend to cooperate with others they like the most. However, collaborating with people from different fields can solve complex problems, such as patents across scientific fields. Furthermore, when this diversity helps to make more dispersed networks more connected, it will produce better scientific outcomes, such as high-quality papers in journals or conferences with more significant impact and higher citation rates. Wang et al. proposed a graph model based on time-constrained probability to identify the consultant-consultant relationship\cite{DBLP:conf/kdd/WangHJTZYG10}.

Analysis based on social networks, Newman et al. used the method of social network analysis to study the macro and micro characteristics of the large-scale academic cooperation network\cite{Newman404}, such as node degree, node centrality, network clustering coefficient, etc., academic cooperation research has aroused people's interest again. Following up on Newman's work in 2001, Barabasi et al. investigated the dynamics and evolution of author collaboration networks\cite{BARABASI2002590}. Since then, authors' collaborative networks have been extensively studied in various ways in the natural and social sciences. Kempe et al. also used author collaboration networks to identify the most influential authors\cite{DBLP:conf/icalp/KempeKT05}. Liu et al. attempted to assess the identity of authors in a particular field by looking at their collaborative networks, thereby strengthening ties with the community by identifying the most influential researchers\cite{6690044,LIU201529}.

In addition to identifying the most influential authors in academic networks, the task of educational relationship mining based on academic author networks also includes modelling cooperation patterns\cite{DBLP:journals/corr/Petersen15g,DBLP:journals/scientometrics/WangYBKX17},recognition of academic relationship\cite{DBLP:conf/kdd/WangHJTZYG10}, community detection\cite{DBLP:journals/tcss/YuXL19}, and collaborator recommendations\cite{DBLP:conf/www/KongJBWX17,DBLP:2012RFH,DBLP:journals/tcss/WangLYKX19,DBLP:journals/tetc/XiaCWLY14}.

\subsection{Research on representation learning in complex networks}

The 5G IoT social data boom is not only revolutionizing our daily lives, but it's also generating a considerable amount of 5G-related paper data. In addition, the feature matrix of the academic cooperation network formed by 5G papers is mostly non-negative.  Therefore, 5G Academic Social network analysis based on non-negative matrix factorization has various application scenarios.  Matrix decomposition-based graph embedding models usually express graphs as matrices, such as the adjacency matrix or approximation matrix of a graph, and decompose the matrices by factoring to obtain graph embeddings or node embeddings that preserve the structural features of the graph\cite{DBLP:journals/kbs/GoyalF18}. There are two types of graph embedding based on matrix factorization: one factorises the Laplacian feature map of the graph, and the other factorises the approximate node matrix directly. The key to factor the Laplacian eigenmaps is to make the final embedded graph feature explain the similarity of paired nodes. Therefore, the greater the embedding distance between two nodes with a more remarkable parallel, the greater the penalty will be to preserve the similarity characteristics between nodes in the network structure. In the initial study MDS\cite{DBLP:conf/nips/HofmannB94}, the distance between two eigenvectors was calculated as the strength of similarity of two nodes. However, this approach has the disadvantage that it does not take into account the neighbours around the node, i.e. any node in all graphs is considered to be connected to any of the remaining nodes. In order to overcome the shortcomings of the complexity of the problem, LPP\cite{DBLP:conf/nips/HeN03}first starts from the characteristics of data structure $k $ most adjacent graph, where each node only calculates the Euclidean distance of the  $k $ nearest node, i.e. the proximity of the node to  $k $ a neighbouring nodes, and then uses the proximity of different nodes to constrain the formation of additional penalties. More advanced models have recently been designed in recent years. For example, AGLPP\cite{DBLP:journals/ijon/JiangFWHH16}realized the rapid improvement of model LPP efficiency by introducing an anchor diagram.

The incorporation of deep learning ideas into non-negative matrix factorisation has become a popular area of research.  NMF can be used to learn data representation by decomposing multivariate data into the product of linear combination basis and auxiliary matrix. By adding additional constraints and penalty terms to induce the thinness\cite{DBLP:conf/nnsp/Hoyer02}and the NSNMF\cite{DBLP:journals/pami/Pascual-MontanoCKLP06}after adding smoothness, etc., an extension of NMF is formed to improve its performance. However, these methods have two inherent problems: first, they assume that the input data can be reconstructed linearly from the base; second, they are all single-layer structural learning, and only basic low-level features can be obtained from the original data.

There has been some exploration in deep matrix decomposition to extract high level non-linear features in network structures\cite{DBLP:journals/corr/GuoZ17ab,DBLP:journals/ijon/SongKLL15,DBLP:conf/icml/TrigeorgisBZS14,DBLP:journals/access/YuZCX18}. Their general idea is to decompose a matrix layer into multiple layers, hoping to get a hierarchical mapping. Trigeorgis et al. proposed a multi-layer semi-NMF model with a complete depth architecture for automatic learning of attribute hierarchies to facilitate clustering tasks\cite{DBLP:conf/icml/TrigeorgisBZS14}. Song et al. proposed a multi-layer NMF structure for classification tasks, in which a non-smooth NMF was used to solve the typical NMF in each layer\cite{DBLP:journals/ijon/SongKLL15}. Then, a sparse depth NMF model is proposed, and Nesterov accelerated gradient descent algorithm\cite{DBLP:journals/corr/GuoZ17ab} is successfully applied to the light structure of data objects. Recently, Yu et al. proposed a deep non-smooth NMF architecture to learn partial and hierarchical attributes\cite{DBLP:journals/access/YuZCX18}. However, all of these models consist only of decoder components. The LNLM model proposed in this paper is autoencoder-like with an objective function that combines an encoder and a decoder. The low-dimensional feature matrix hidden in the original network is transformed by the low-dimensional feature matrix hidden in the intermediate layers.  For each layer of the encoder, the similarities between nodes at different granularity levels are explained. The decoder components seek to learn from the hierarchical mapping of the encoder components to effectively fuse and reconstruct the multi-feature matrix of the original network.

\begin{table}[h]
	\caption{
		Notations}\label{notation}
	\vspace{0.5em}\centering
	\begin{tabular}{cc}
		\toprule[1.5pt]
		Symbol &Definition\\
		\midrule[1pt]
		$G=(V,E)$ & $\textbf{$\mathbb{A}$}$ graph $G$ with node set $V$ and edge set $E$ \\
		$\textbf{$\mathbb{A}$}\in \textbf{$\mathbb{R}$}^{n\times n}$ & Adjacency matrix of graph $G$\\
		$\textbf{$\mathbb{B}$}\in \textbf{$\mathbb{R}$}^{n \times d}$& Low-order eigenmatrix of nodes \\
		$\textbf{$\mathbb{V}$}\in \textbf{$\mathbb{R}$}^{n \times k}$& Representation matrix of the $k$-dimension nodes \\
		$\textbf{$\mathbb{Z}$}\in \textbf{$\mathbb{R}$}^{n \times m}$& Local eigenmatrix of nodes \\		
		$vol(G)=\sum_{i} \sum_j \textbf{$\mathbb{A}$}_{\textbf{ij}}$& Capacity of Graph $G$\\
		$T$& Sliding Window Size\\
		\bottomrule[1.5pt]
	\end{tabular}
	\vspace{\baselineskip}
\end{table}

\section{Low-order networks represent learning to model}
\subsection{Model and its solution}
Given an undirected network, $G=(V,E)$, including $n $ node and $E $ side, $V=\{v_1,v_2,...,v_n\}$ represents a set of nodes, and   $E$ represents a collection of edges between nodes. $G$ can be represented by an adjacency matrix $\textbf{$\mathbb{A}$}\in {\textbf{$\mathbb{R}$}^{n\times n}}$, denoted $\textbf{$\mathbb{A}$}(i,:)$ representing the connections between node $i$ and other nodes in $V$  in the connection between the other nodes. For a weighted network, if there is an edge between node $i$ and node $j$, otherwise $\textbf{$\mathbb{A}$}_{\textbf{ij}}=0$. Since the network $G$ is undirected,  $\textbf{$\mathbb{A}$}$ is a symmetric matrix, i.e. $\textbf{$\mathbb{A}$}_{\textbf{ij}}=\textbf{$\mathbb{A}$}_{\textbf{ji}}$. Throughout this paper, the matrix is shown using bold capital characters. The matrix $\textbf{$\mathbb{B}$}\in \textbf{$\mathbb{R}$}^{n \times d}$ preserves the low-order structural features of the nodes in the network, where  $d$ is the dimension of the feature. $B(i,:)$ describes the structural characteristics of the node  $i$. The purpose is to learn the representation of the node $\textbf{$\mathbb{V}$}\in \textbf{$\mathbb{R}$}^{n\times k}(k\le n)$, where $k$ is the dimension of the representation. The table \ref{notation} shows the notation used in this paper.

In the design of the LNLM model, this paper mainly considers two essential parts, a local strCapacity of Graph Guctural feature encoder and a lower order feature encoder. The adjacency matrix  $\textbf{$\mathbb{A}$}$ retains most of the network topology characteristics and directly represents the first-order similarity. The model first captures the coarse-grained structure feature matrix $\textbf{$\mathbb{X}$}$ from the original network by decomposing the adjacency matrix $\textbf{$\mathbb{A}$}$.
\subsection{Local structural feature extraction}
In the real world, information networks often lose a lot of information, and many nodes in the adjacency matrix that are not directly connected also have high similarity in nature. In addition, the adjacency matrix retains directly connected edges that represent the first-order similarity of the nodes. In particular, they reveal that the adjacency matrix preserves the topology of the network. Therefore, the adjacency matrix  $\textbf{$\mathbb{A}$}$ is decomposed to capture the topology of the network. In this paper, the NMF method is adopted to maintain local structure in low-dimensional space and minimize the following objective functions to the maximum extent:
\begin{equation}
\centering
\begin{aligned}
\underset{\textbf{$\mathbb{Z}$}\ge 0}{min} \ \lVert \textbf{$\mathbb{A}$}-\textbf{$\mathbb{ZZ}$}^\textbf{$\mathbb{T}$}\rVert^{2}_{F} \quad
\end{aligned}
\end{equation}
Among them $\textbf{$\mathbb{Z}$}\in \textbf{$\mathbb{R}$} ^{n \times m}$ is the local eigenmatrix, $m$ it's the spatial dimension.

Because embedding the matrix  $\textbf{$\mathbb{Z}$}$  loses a lot of crucial information, lower-order features are integrated to mutually enhance the learning of $\textbf{$\mathbb{Z}$}$ at the same time, by decomposing the local eigenmatrix $\textbf{$\mathbb{Z}$}$ to obtain the low-dimensional representation of nodes that retain local and community structures, the following objective functions are minimized:
\begin{equation}
\underset{\textbf{$\mathbb{Z}$} \ge 0,\textbf{$\mathbb{V}$} \ge 0, \textbf{$\mathbb{U}$} \ge 0}{min} \ \lVert \textbf{$\mathbb{Z}$}-\textbf{$\mathbb{VU}$} \rVert^{2}_{F} \quad 
\end{equation}

Among them $\textbf{$\mathbb{V}$}\in \textbf{$\mathbb{R}$}^{n \times k}$ it's an embedded matrix, $\textbf{$\mathbb{U}$}\in \textbf{$\mathbb{R}$}^{k \times m}$ is the auxiliary matrix, $k$ is the embedded dimension.

\subsection{Lower order structural feature extraction}

In this paper, lower-order features of nodes are extracted based on a random walk. Zhang et al. proved that the model based on a random walk and graph hop is considered to be of the matrix decomposition, a closed-form and verify the effectiveness of them for the conventional network mining tasks\cite{DBLP:conf/wsdm/QiuDMLWT18}. The matrix representation of implicit approximation and factorization is as follows:
\begin{equation}
\textbf{$\mathbb{M}$}=log(vol(G)(\frac{1}{T}\sum_{r=1}^{T}{\textbf{$\mathbb{D}$}^{-1}\textbf{$\mathbb{A}$}}^r\textbf{$\mathbb{D}$}^{-1}))-log(b)
\end{equation}

Where $vol(G)$ is the capacity of the graph $G$,  $vol(G)=\sum_i \sum_j \textbf{$\mathbb{A}$}_{\textbf{ij}}$.  $\textbf{$\mathbb{D}$}$ is degree matrix of graph $G$, $\textbf{$\mathbb{D}$}=diag(d_1,d_2,...,d_{\lvert V \lvert})$. $T$ is the sliding window, and $b$ is the negative sampling number in the skip-gram model. Truncated Singular Value Decomposition (SVD) is then performed on the constructed $\textbf{$\mathbb{M}$}$ to capture the eigenmatrix $\textbf{$\mathbb{B}$}$ of the node in the low-dimensional space, that is, $\textbf{$\mathbb{M}$} = \textbf{$\mathbb{BV}$}$. As the window size  $T$ increases, lower order structures can be obtained. Inspired by this, lower order structural features are extracted from the matrix $\textbf{$\mathbb{B}$}$

\begin{equation}
\underset{\textbf{$\mathbb{V}$} \ge 0,\textbf{$\mathbb{H}$} \ge 0}{min} \ \lVert \textbf{$\mathbb{VH}$}-\textbf{$\mathbb{B}$}\rVert^2_F 
\end{equation}

$\textbf{$\mathbb{H}$}\in \textbf{$\mathbb{R}$}^{k\times d}$ is an auxiliary matrix.
\section{Model Building}
\begin{figure}[htbp]
	\centering
    \includegraphics[width=\linewidth]{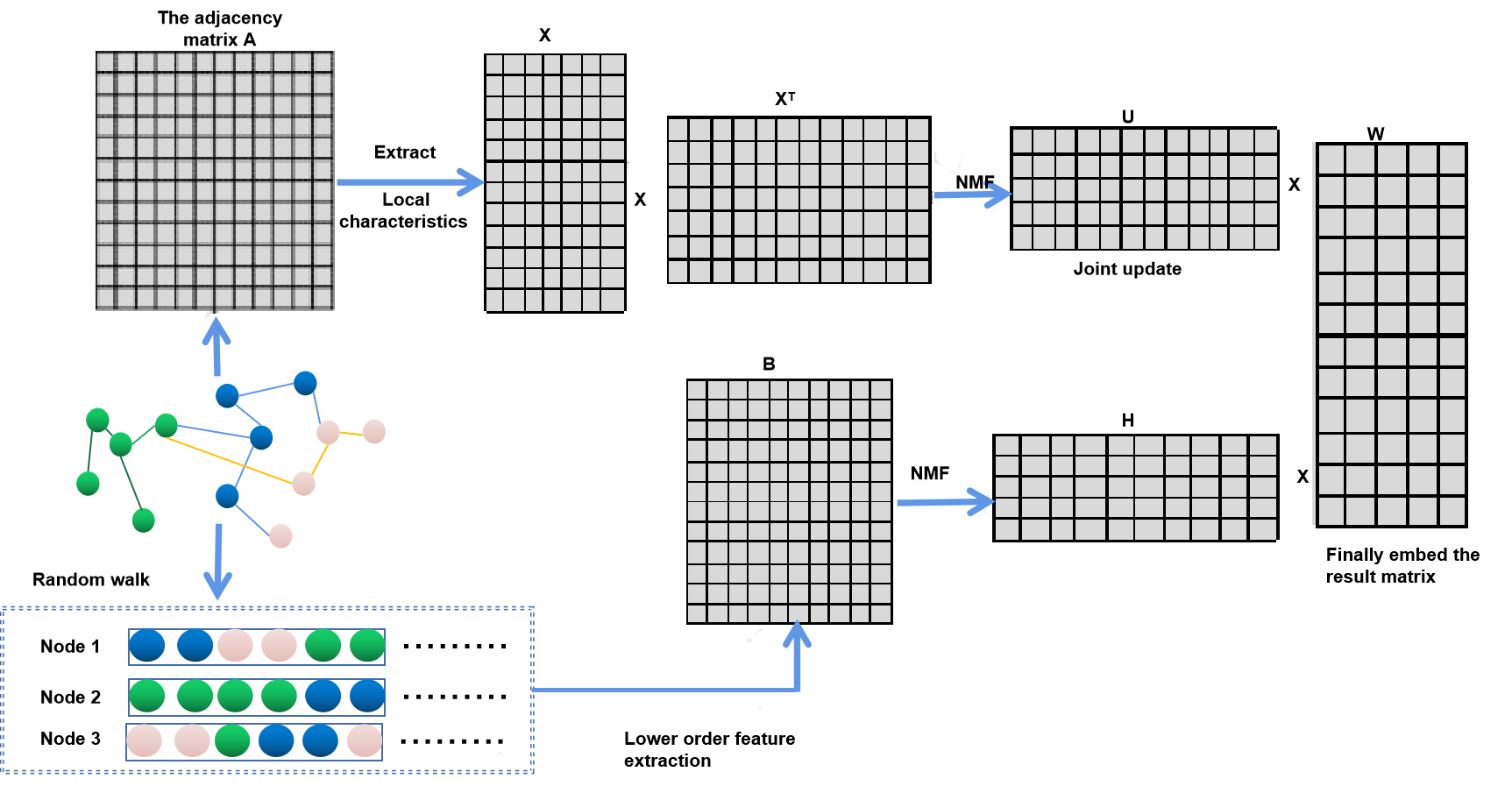}
	\caption{Framework of LNLM model}
	\label{fig1}
	\vspace{\baselineskip} 
\end{figure}

Because the network is very sparse, many node features will be lost only through the first-order similarity of nodes. At the same time, the random walk sequence is generated for each node, and the $k$ order feature is learned by controlling the window size. In addition, the lower order features known from the random walk are integrated into the NMF framework, and the local structures are captured to be represented by learning nodes so that the lower order feature information and regional facilities are retained together. The overall architecture of this model is shown in Figure \ref{fig1}.

After integrating the objective functions (1), (2) and (4), the definition of the final loss function of this model is as follows:
\begin{equation}
\centering
\begin{aligned}
\label{5}
\underset{\textbf{$\mathbb{Z}$} \ge 0, \textbf{$\mathbb{V}$} \ge 0, \textbf{$\mathbb{H}$} \ge 0,\textbf{$\mathbb{U}$} \ge 0}{min}\ \mathcal{L}\! =\!\underset{\textbf{$\mathbb{Z}$} \ge 0, \textbf{$\mathbb{V}$} \ge 0, \textbf{$\mathbb{H}$} \ge 0,\textbf{$\mathbb{U}$} \ge 0}{min} \ \lVert \textbf{$\mathbb{A}$}-\textbf{$\mathbb{ZZ}$}^\textbf{$\mathbb{T}$}\rVert^{2}_{F}+\alpha \lVert \textbf{$\mathbb{Z}$}-\textbf{$\mathbb{VU}$} \rVert^{2}_{F} 
+ \beta \lVert \textbf{$\mathbb{VH}$}-\textbf{$\mathbb{B}$}\rVert^2_F +\\ \gamma (\lVert \textbf{$\mathbb{U}$}\rVert^2_F+ \lVert \textbf{$\mathbb{H}$}\rVert^2_F) 
\end{aligned}
\end{equation}

Where $\alpha $,  $\beta$,  $\gamma$ are positive parameters used to adjust the contribution of the corresponding item.
\subsection{Model optimization} 
Since the loss function in the formula (\ref{5}) is not convex, the derivative cannot be used to calculate the optimal solution. In this paper, the loss function is divided into four subproblems, and the four parameter matrix $(\textbf{$\mathbb{Z}$},\textbf{$\mathbb{V}$},\textbf{$\mathbb{H}$},\textbf{$\mathbb{U}$})$ is optimized respectively.

Then, using the Majorization - Minimization framework\cite{doi:10.1198/0003130042836}each optimal local solution to the problem is then updated using the majorisation-minimisation framework. The update strategy adopted is alternate optimization, that is, when one matrix is updated, the other three matrices are fixed. The algorithm \ref{algorithm1} shows the pseudocode for the optimization process. The specific formula is as follows:

The $\textbf{$\mathbb{Z}$}$-related loss function is as follows: 
\begin{equation}
\label{6}
\underset{\textbf{$\mathbb{Z}$} \ge 0}{min} \ \lVert \textbf{$\mathbb{A}$}-\textbf{$\mathbb{ZZ}$}^\textbf{$\mathbb{T}$}\rVert^{2}_{F}+\alpha \lVert \textbf{$\mathbb{Z}$}-\textbf{$\mathbb{VU}$} \rVert^{2}_{F}\quad
\end{equation}	

Here, $\textbf{$\mathbb{Z}$}$ is a non-negative matrix, so the Lagrange multiplier matrix $\Theta$ is introduced to obtain the following equivalent function:
\begin{equation}
\centering
\begin{aligned}
\label{7}
L(\textbf{$\mathbb{Z}$})=tr(\textbf{$\mathbb{AA}$}^{\textbf{$\mathbb{T}$}}-\textbf{$\mathbb{AZZ}$}^{\textbf{$\mathbb{T}$}}-\textbf{$\mathbb{ZZ}$}^{\textbf{$\mathbb{T}$}}\textbf{$\mathbb{A}$}^{\textbf{$\mathbb{T}$}}-\textbf{$\mathbb{ZZ}$}^{\textbf{$\mathbb{T}$}}\textbf{$\mathbb{ZZ}$}^{\textbf{$\mathbb{T}$}})\\
+atr(\textbf{$\mathbb{ZZ}$}^{\textbf{$\mathbb{T}$}}-\textbf{$\mathbb{ZU}$}^{\textbf{$\mathbb{T}$}}\textbf{$\mathbb{V}$}^{\textbf{$\mathbb{T}$}}-\textbf{$\mathbb{VUZ}$}^{\textbf{$\mathbb{T}$}}+\textbf{$\mathbb{VUU}$}^{\textbf{$\mathbb{T}$}}\textbf{$\mathbb{V}$}^{\textbf{$\mathbb{T}$}})
+tr(\Theta \textbf{$\mathbb{Z}$}^{\textbf{$\mathbb{T}$}})
\end{aligned}
\end{equation}

Set the value of formula (\ref{7}) to 0, that is $L(\textbf{$\mathbb{Z}$})=0$:
\begin{equation}
\label{8}
\Theta=4\textbf{$\mathbb{AZ}$}-4\textbf{$\mathbb{ZZ}$}^{\textbf{$\mathbb{T}$}}\textbf{$\mathbb{Z}$}-2\alpha \textbf{$\mathbb{Z}$}+2\alpha \textbf{$\mathbb{VU}$}
\end{equation}

Following the Karush-Kuhn-Tucker (KKT) condition on the non-negative property of   $\textbf{$\mathbb{Z}$}$, the following equation is obtained:
\begin{equation}
\label{9}
(4\textbf{$\mathbb{AZ}$}-4\textbf{$\mathbb{ZZ}$}^{\textbf{$\mathbb{T}$}}\textbf{$\mathbb{Z}$}-2\alpha \textbf{$\mathbb{Z}$}+2\alpha \textbf{$\mathbb{VU}$})\cdot{\textbf{$\mathbb{Z}$}}=\Theta \textbf{$\mathbb{Z}$}=0
\end{equation}

Initialization and updates based on $\textbf{$\mathbb{Z}$}$  are as follows
\begin{equation}\label{10}
\textbf{$\mathbb{Z}$} \leftarrow \textbf{$\mathbb{Z}$} \odot (\frac{2\textbf{$\mathbb{AZ}$}+\alpha \textbf{$\mathbb{VU}$}}{2\textbf{$\mathbb{ZZ}$}^{\textbf{$\mathbb{T}$}}\textbf{$\mathbb{Z}$}+\alpha \textbf{$\mathbb{Z}$}})
\end{equation}
Where $\odot$ represents multiplying matrices

On the optimization of V-subproblem,
when updating $\textbf{$\mathbb{V}$}$, the fixed arguments  $\textbf{$\mathbb{Z}$}$, $\textbf{$\mathbb{H}$}$, $\textbf{$\mathbb{U}$}$resolve the following target function:
\begin{equation}
\underset{\textbf{$\mathbb{V}$} \ge 0}{min} \ \alpha \lVert \textbf{$\mathbb{Z}$}-\textbf{$\mathbb{VU}$} \rVert^{2}_{F} 
+ \beta \lVert \textbf{$\mathbb{VH}$}-\textbf{$\mathbb{B}$}\rVert^2_F \quad 
\end{equation}

Similar to $\textbf{$\mathbb{Z}$}$, the update rules that define $\textbf{$\mathbb{V}$}$ are as follows:
\begin{equation}\label{12}
\textbf{$\mathbb{V}$} \leftarrow \textbf{$\mathbb{V}$} \odot (\frac{\alpha \textbf{$\mathbb{ZU}$}^{\textbf{$\mathbb{T}$}}+\beta \textbf{$\mathbb{BHZ}$}^{\textbf{$\mathbb{T}$}}}{\alpha \textbf{$\mathbb{VUU}$}^{\textbf{$\mathbb{T}$}}+\beta \textbf{$\mathbb{VHH}$}^{\textbf{$\mathbb{T}$}}})
\end{equation}

On the optimization of H-subproblem,
when updating $\textbf{$\mathbb{H}$}$ the fixed parameters $\textbf{$\mathbb{Z}$}$, $\textbf{$\mathbb{V}$}$, $\textbf{$\mathbb{U}$}$, will have the following objective functions:
\begin{equation}
\underset{\textbf{$\mathbb{H}$} \ge 0}{min} \ \alpha \ \lVert \textbf{$\mathbb{VH}$}-\textbf{$\mathbb{B}$} \rVert^{2}_{F} 
+ \gamma \lVert \textbf{$\mathbb{H}$} \rVert^2_F 
\end{equation}

Similarly to  $\textbf{$\mathbb{Z}$}$, the update rule that defines $\textbf{$\mathbb{H}$}$ is as follows:
\begin{equation}\label{14}
\textbf{$\mathbb{H}$} \leftarrow \textbf{$\mathbb{H}$} \odot (\frac{\textbf{$\mathbb{V}$}^{\textbf{$\mathbb{T}$}}\textbf{$\mathbb{B}$}}{\beta \textbf{$\mathbb{V}$}^{\textbf{$\mathbb{T}$}}\textbf{$\mathbb{VH}$}+\gamma \textbf{$\mathbb{H}$}})
\end{equation}

On the optimization of U-subproblem, 
when updating $\textbf{$\mathbb{U}$}$ the fixed parameters $\textbf{$\mathbb{Z}$}$, $\textbf{$\mathbb{V}$}$, $\textbf{$\mathbb{H}$}$, will have the following target functions:
\begin{equation}
\underset{\textbf{$\mathbb{U}$} \ge 0}{min}\ \beta \ \lVert \textbf{$\mathbb{Z}$}-\textbf{$\mathbb{VU}$} \rVert^{2}_{F} 
+ \gamma \lVert \textbf{$\mathbb{U}$} \rVert^2_F \quad 
\end{equation}

Similarly, similar to the optimization calculation of $\textbf{$\mathbb{Z}$}$, the update rule that defines $\textbf{$\mathbb{U}$}$ is as follows:
\begin{equation}\label{16}
\textbf{$\mathbb{U}$} \leftarrow \textbf{$\mathbb{U}$} \odot (\frac{\alpha \textbf{$\mathbb{V}$}^{\textbf{$\mathbb{T}$}}\textbf{$\mathbb{Z}$}}{\alpha \textbf{$\mathbb{V}$}^{\textbf{$\mathbb{T}$}}\textbf{$\mathbb{VU}$}+\gamma \textbf{$\mathbb{U}$}})
\end{equation}

An algorithm \ref{algorithm1} describes the optimization process of the method. The optimized input data includes network $G$, lower order eigenmatrix  $\textbf{$\mathbb{B}$}$, embedding dimension $k$, convergence coefficient $\delta$ and balance parameters $\alpha$, $\beta$, $\gamma$. First of all, with uniformly distributed random initialization $\textbf{$\mathbb{Z}$}$, $\textbf{$\mathbb{V}$}$, $\textbf{$\mathbb{U}$}$, $\textbf{$\mathbb{H}$}$. Then, iteratively update  $\textbf{$\mathbb{Z}$}$, $\textbf{$\mathbb{V}$}$, $\textbf{$\mathbb{U}$}$, $\textbf{$\mathbb{H}$}$ until convergence occurs. The output is the embedded matrix $\textbf{$\mathbb{V}$}$ for all nodes in the network. The node representation learned from the model in this paper can obtain lower order features and local structures.

\begin{algorithm}[]
	\label{algorithm1}
	\caption{LNLM model optimization}%
	\LinesNumbered %
	\hspace*{0.02in}
    \KwIn{Network $G$, lower order eigenmatrix $\textbf{$\mathbb{A}$}$, embedded dimension $d$, convergence coefficient  $\delta$, balance parameter $\alpha $, $\beta$, $ \gamma$}%
	\KwOut{$\textbf{$\mathbb{Z}$}\in \textbf{$\mathbb{R}$}^{n \times m}$, $\textbf{$\mathbb{V}$} \in \textbf{$\mathbb{R}$}^{n \times k}$, $\textbf{$\mathbb{U}$} \in \textbf{$\mathbb{R}$}^{k \times m}$}%
	\textbf{Initial:}$\textbf{$\mathbb{Z}$}$, $\textbf{$\mathbb{V}$}$, $\textbf{$\mathbb{U}$}$, $\textbf{$\mathbb{H}$}$\\ 
	\While{not conv}{
		\If{$\frac{\mathcal{L}^i-\mathcal{L}^{i-1}}{\mathcal{L}^{i-1}}< \delta $}{
			(\ref{10}) with the formula $\textbf{$\mathbb{Z}$}$ \\
			(\ref{12}) with the formula $\textbf{$\mathbb{V}$}$\\
			(\ref{14}) with the formula $\textbf{$\mathbb{H}$}$\\
			(\ref{16}) with the formula $\textbf{$\mathbb{U}$}$\\
			Calculate the loss function (\ref{5}) using the formula  $\mathcal{L}$\\
			$i \leftarrow i+1$
		}
		\Else{
			conv$\leftarrow\textbf{true}$ \\
		}
	}
	\Return $\textbf{$\mathbb{Z}$}$, $\textbf{$\mathbb{V}$}$, $\textbf{$\mathbb{U}$}$, $\textbf{$\mathbb{H}$}$
\end{algorithm}

\subsection{Model complexity }
The entire computational complexity of the model depends on the matrix multiplication in the update rule, so given two matrices $\textbf{$\mathbb{A}$}$ and $\textbf{$\mathbb{B}$}$,Where $\textbf{$\mathbb{A}$}\in \textbf{$\mathbb{R}$}^{m\times r}$ and $\textbf{$\mathbb{B}$}\in \textbf{$\mathbb{R}$}^{r\times n}$, $\textbf{$\mathbb{AB}$}$ computational complexity is  $O(mrn)$. Based on this, the computational complexity of the four update formulas in the algorithm \ref{algorithm1} is $O(n^2m+nkm+nm^2)$, $O(nmk+ndk+2nk^2+mk^2+ dk^2)$, $O(nkd+nk^2+dk^2)$, $O(nkm+nk^2+mk^2)$ . Since $m$, $k$, $d$ can be considered as input constants, and $m\le n$, $k \ll n$, $d\ll n$, the computational complexity is  $O(n^2+nm^2+nkm+nkd)$. In fact, most networks are very sparse, so only non-zero values are evaluated in matrix multiplication. Based on this, the calculation is simplified to $O(ne+nm^2+ nkm+nkd)$, where $e$ is the number of edges in the network.  $\textbf{$\mathbb{Z}$}\in \textbf{$\mathbb{R}$}^{n \times m}$, $\textbf{$\mathbb{V}$}\in \textbf{$\mathbb{R}$}^{n\times k}$, $\textbf{$\mathbb{U}$}\in \textbf{$\mathbb{R}$}^{k\times m}$, $\textbf{$\mathbb{H}$}\in \textbf{$\mathbb{R}$}^{k\times d}$ is the parameter matrix, so the space complexity is  $O(nm+nk+km+kd)$. Because $k$, $m$, and $d$ are less than $n$, the spatial calculation is simplified to $O(n)$. It can be found that for most NMF-based algorithms, the complexity of the model is on the same order of magnitude.

To effectively illustrate the temporal complexity of the model, the low-dimensional embedding of nodes for several datasets of different sizes was learned, and the time was calculated and then displayed in the table \ref{time_ca}. The results show that the size of the network increases exponentially, and the computing time also increases exponentially.
\begin{table}[h]
	\caption{Computation time on different datasets}\label{time_ca}
	\vspace{0.5em}\centering
	\begin{tabular}{ccccc}
		\toprule[1.5pt]
		Dataset &OAG &Wikipedia&Polblog &Hep-ph\\
		\midrule[1pt]
		Number of nodes& 13890 & 4777&10312&80513\\
		Time &17.113s&0.683s&12.152s&77.330s\\	
		\bottomrule[1.5pt]
	\end{tabular}
	\vspace{\baselineskip}
\end{table}

\section{Experimental Verification}

In this section, extensive experiments will be conducted on multi-tag node classification, clustering, link prediction and visualization tasks to evaluate the model's effectiveness. First, the data sets and experimental Settings used in this paper are described. In order to prove the validity of the model, it is compared with the latest methods in extensive experiments. Finally, the sensitivity of using parameters is analyzed. All the experiments were run on computers configured with a Windows 8 64-bit operating system, a 3.10 GHz CPU and 256 GB RAM. The detailed comparison algorithm is as follows:

\subsection{Datasets}
This section focuses on four widely used network data sets for multi-label node classification tasks and four real networks, including 5G academic social networks with basic fact data sets for clustering and link prediction. The statistical characteristics of these datasets are shown in table \ref{data_set_1}, which is described in detail below. 
\begin{itemize}
	\item Wikipedia:This is the word co-occurrence network on Wikipedia. Class tags are part of speech (POS) tags inferred by Stanford PoS Tagger\cite{DBLP:conf/naacl/ToutanovaKMS03}.
	\item Polblog\cite{adamic2005political}:PolBlog, a social network whose nodes represent the blogs of US politicians, has an advantage if their blogs have Weblinks. The tag indicates the type of politician.
	\item Livejournal\cite{DBLP:journals/kais/YangL15}: LiveJournal is an online social network data set whose nodes represent bloggers and have edges between two nodes if they are friends. Divide the bloggers into groups based on their friendships and use these groups as tags.
	\item Orkut\cite{DBLP:journals/kais/YangL15}: Orkut is an online dating network that uses nodes as users and builds links between nodes based on their friends. The Web has several clearly labelled communities, including student communities, activities, interest-based groups, and school teams.
	\item GRQC,Hep-th,Hep-ph\cite{DBLP:journals/tkdd/LeskovecKF07}: This is a collaborative network of three authors in general relativity and quantum cosmology, theory of high energy physics, and phenomenology of high energy physics extracted from arXiv. In this network, vertices represent authors, and edges represent authors who have co-authored a scientific paper in arXiv.
	\item Open Academic Graph (OAG)\cite{chaudhuri2012spectral}: This is an undirected author collaboration network constructed from a publicly available academic chart indexed by Microsoft Academic and American Miner websites. The network contains 67,768,244 authors and 895,368,962 collaborative advantages. Vertex tags are defined as each author's top research area, such as computer science, physics, psychology, etc. There are 19 different fields (labels) in total, and the author can publish in multiple fields, which makes the corresponding vertices have multiple labels.
	\item Academic Social Network(ASN)\cite{wan2019aminer}:Data included 2,092,356 papers, with 8,024,869 citations, 1,712,433 authors and 4,258,615 co-authors. Among them, the number of nodes related to 5G is 2407, and the number of edges is 1836. In the past 10 years, the top  highly relevant to 5G each year are 6143  in total, cited 100572 times by 2635 authors and 6316 co-authors. According to the domestic, foreign papers, and related patents on the line to label and the author cooperation network of different types of research, to brand.
	
\end{itemize}
\begin{table}[h]
	\caption{Dataset Statistics}\label{data_set_1}
	\vspace{0.5em}\centering
	\begin{tabular}{ccccc}
		\toprule[1.5pt]
		Dataset & Node Number $\lvert V\rvert$ & Edge Number $\lvert E\rvert$& Category Number $k$ & Multiple Tags \\
		\midrule[1pt]
		Wikipedia&4777&18412&40&yes\\
		Polblog&1490&16627&2&no\\
		Livejournal&11118&396461&26&no\\
		Orkut&998&23050&6&no\\	
		GRQC&4158&13422&42&no\\
		Hep-th&8638&24806&5&yes\\
		Hep-ph&11204&57619&38&yes\\
		OAG&13890&86784&19&yes\\
		ASN&10407&14156&9&yes\\
		\bottomrule[1.5pt]
	\end{tabular}
	\vspace{\baselineskip}
\end{table}
\subsection{Baseline methods}
This section highlights the comparison of three and five state-of-the-art network embedding methods based on NMF. The details are as follows:
\begin{itemize}

	\item M-NMF\cite{DBLP:conf/aaai/WangCWP0Y17}: M-NMF unifies the community structure characteristics and the 2-hop neighbourhood relations of nodes in the NMF framework to learn the embedding of nodes in the network structure. The model uses nodes to represent the consistent relationship with the network community structure. 
	It uses an auxiliary community to represent the matrix to connect the local features (first-order similarity and the community structure features in the network structure) and uses the optimization formula to optimize them jointly. The embedded dimension in the experiment is 128, and other parameters in the investigation are set according to the original paper.
	\item NetMF\cite{DBLP:conf/wsdm/QiuDMLWT18}: NETMF proved that models using negative samplings, such as DEEP WALK, PTE and LINE, can be decomposed into closed matrices and confirmed that it is superior to DEEP WALK and LINE in conventional network analysis and mining tasks.
	\item AROPE\cite{DBLP:conf/kdd/ZhangCWPY018}: The AROPE based SVD framework moves the embedded vectors across any order and reveals the intrinsic relationships between them to learn any high order proximity of nodes.
	\item DeepWalk\cite{DBLP:conf/kdd/PerozziAS14}: Deepwalk generates a random path for each node and treats the path of these nodes as a sentence in a language model. It then uses the skip-gram model to learn to embed vectors. In the experiment, the parameters are consistent with those in the original paper.
	\item Node2vec\cite{DBLP:conf/kdd/GroverL16}: Node2vec extends DeepWalk by using biased random walk. It introduces two offset parameters $p$ and $q$ to optimize the random walk. All parameters are default Settings.
	\item LINE\cite{DBLP:conf/www/TangQWZYM15}: LINE learns the embedding of nodes by defining two loss functions to preserve the first-order and second-order proximity, respectively. This article uses the default parameter settings, but the negative ratio is 5.
	\item  GAE\cite{DBLP:journals/corr/KipfW16a}:GAE is based on a variational autoencoder and has the same convolution architecture as GCN. This method has strong competitiveness in link prediction tasks in citation networks.
	\item SDNE\cite{DBLP:conf/kdd/WangC016}:SDNE uses a deep autoencoder with a semi-supervised architecture to optimize the first-order and second-order similarity of nodes simultaneously and uses explicit objective functions to clarify how to preserve the network structure. The parameters in the experiment are consistent with those set in the original paper.
\end{itemize}

\subsection{Parameter sensitivity analysis}
In this section, we analyze the effects of LNLM parameters $\alpha$, $\beta$, $\gamma$ and $m$ on the clustering performance on real networks, where $m$ is the dimension of the local structure embedding space. The experimental results show that the clustering performance presents a similar trend in different data sets with the change of parameters. For simplicity, the effects of different parameters are discussed with an OAG dataset as an example.

\begin{figure}[!h]	
	\centering
	\subfigure[$\alpha,\beta$ Relationship between NMI and cluster evaluation index]{
		\centering
		\includegraphics[width=2.1in]{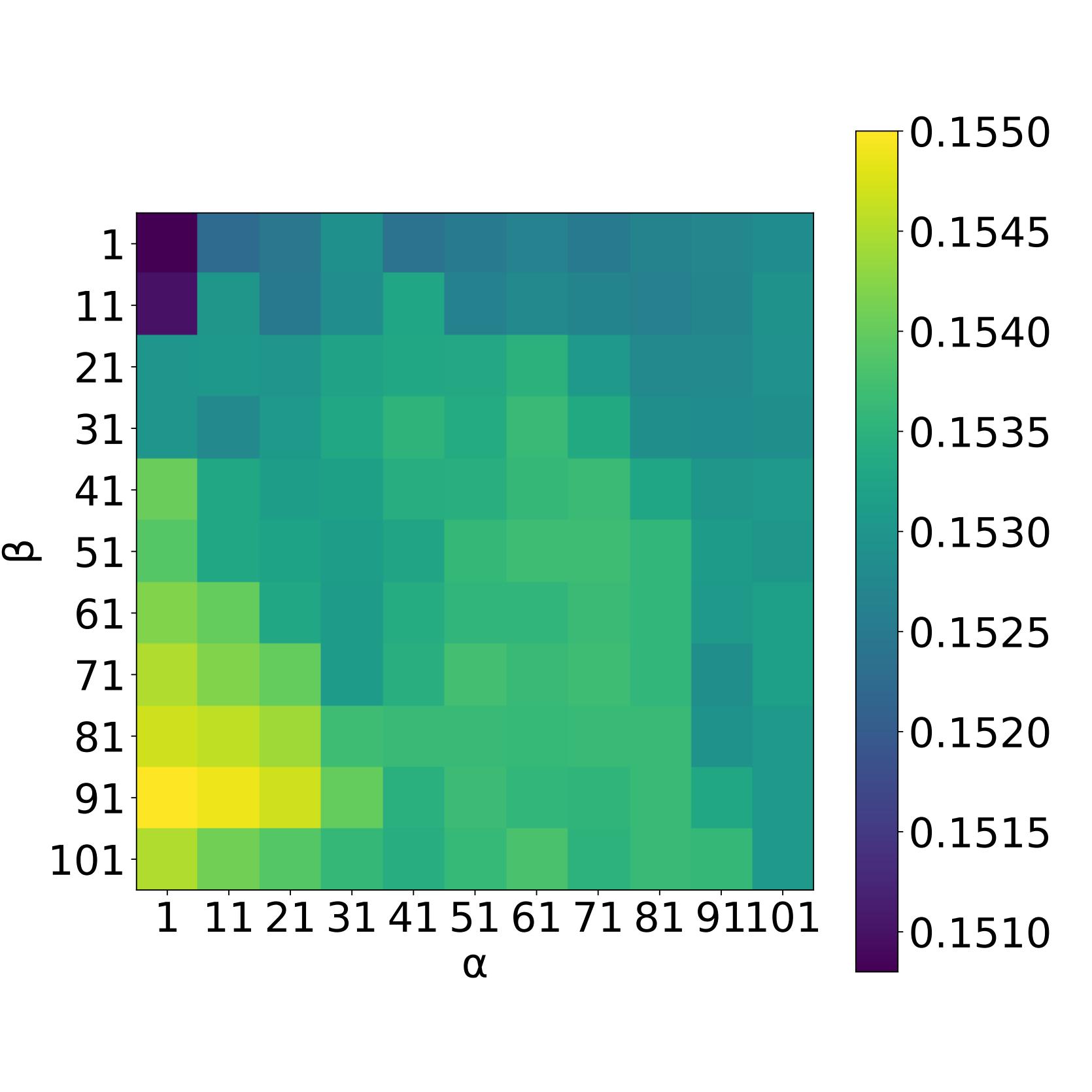}	
	}%
	\quad
	\subfigure[$\alpha,\gamma$ Relationship between NMI and cluster evaluation index]{
		\centering
		\includegraphics[width=2.1in]{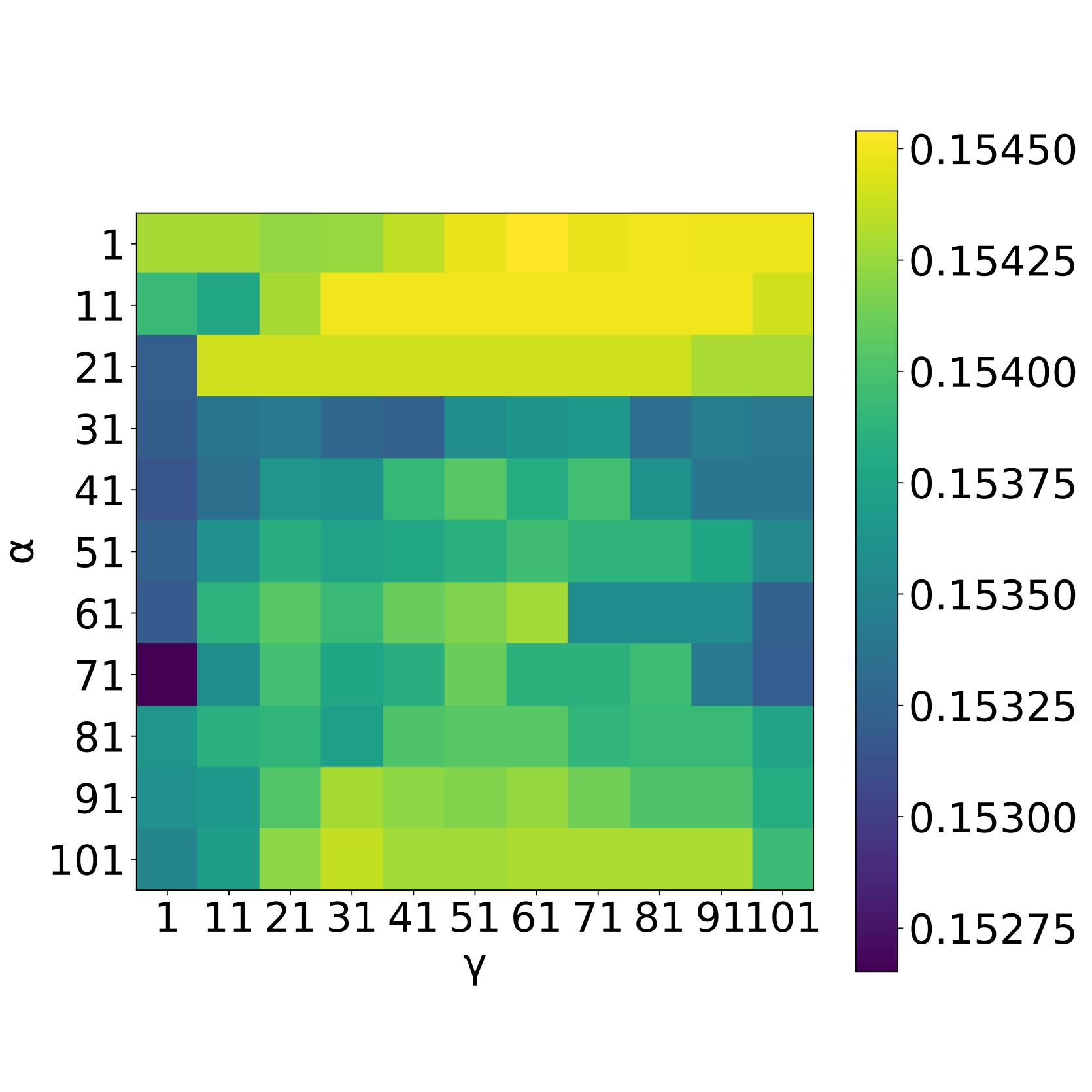}
	}%
	\quad
	\subfigure[$\gamma,\beta$ Relationship between NMI and cluster evaluation index]{
		\centering
		\includegraphics[width=2.1in]{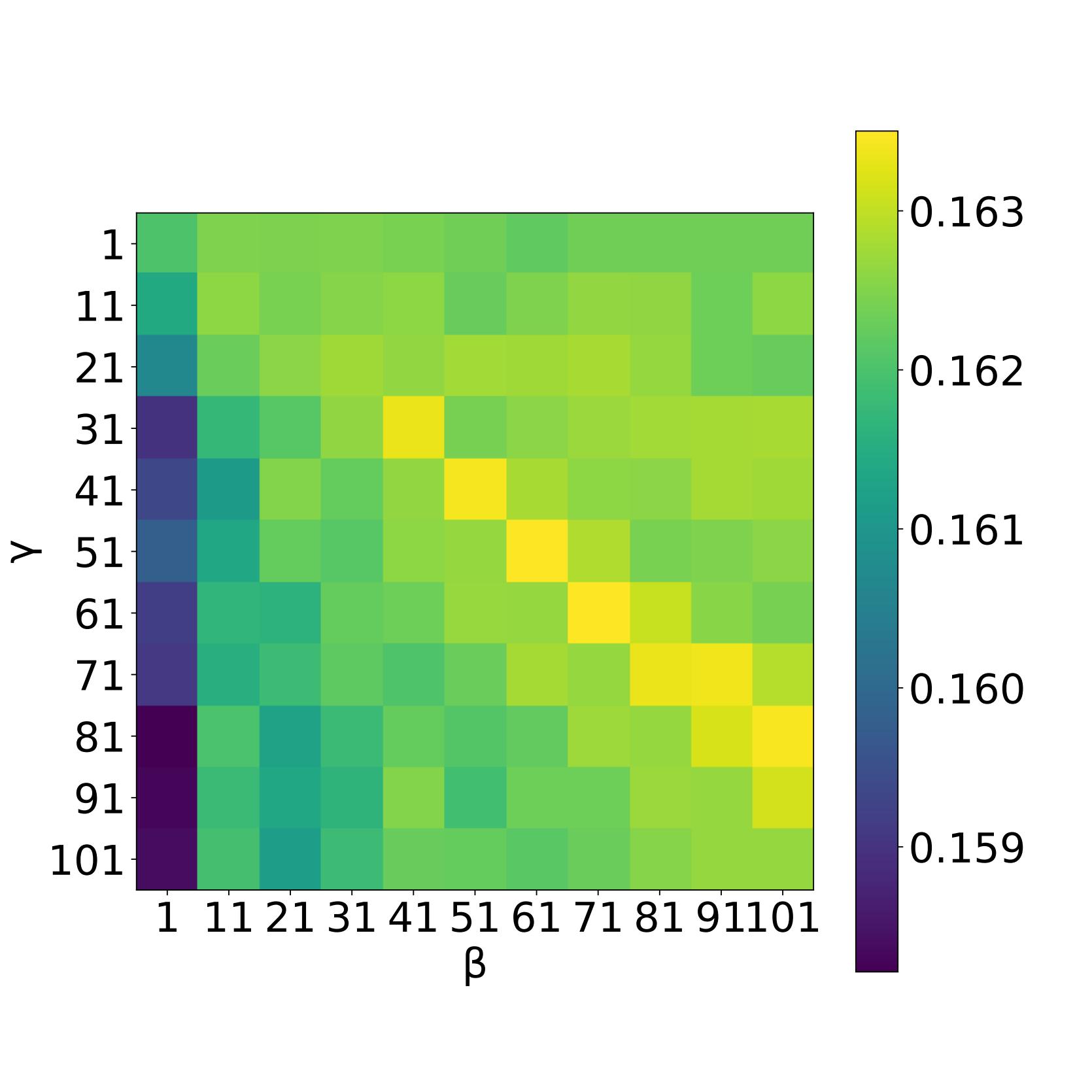}
	}
	\centering
	\caption{Parameter sensitivity in NMI and cluster evaluation index}
	\label{fig2}
\end{figure}

The details of the parameter analysis are shown in Figure \ref{fig2}. The influence of the two parameters on the experimental results is analyzed below. In general, the values of $\alpha$, $\beta$ and $\gamma$ are set based on the data in the following experiment.
Explore the effect of each parameter by changing two parameters while controlling the others. For example, change $\alpha$, $\beta$ and fix $\gamma$ and $m$ to see the effect of $\alpha$, $\beta$. And so on. To be specific, change $m$ from 100, 200, 300, 400 and 500. Figures \ref{fig2}(a)-(c) and Table \ref{t.7} show the performance of NMI as these parameters vary, respectively. In the figure  \ref{fig2}(a), the clustering performance is the worst when both $\alpha$ and   $\beta$ are less than 10. Within certain limits, NMI values tend to be stable as $\alpha$ and $\beta$ increase. The clustering performance is best when $\alpha$ is greater than 50 and $\beta$ is less than 30.

As shown in Figure \ref{fig2}(b), horizontally, NMI does not change much when $\alpha$ in [1,20], indicating that the clustering performance is relatively stable as $\alpha$ is in A and as $\gamma$ is increased. Some range. In the figure \ref{fig2}(c), it is noted that within a specific field, NMI tends to be stable when $\beta$ and $\gamma$ are linearly correlated, while NMI reaches its maximum when $\gamma$ and $\beta$ are in the range [20,101].

The parameters of the LNLM model include three hyperparameters  $\alpha$, $\beta$ and $\gamma$, the local structure size $m$ and the embedded size $k$. In the experiment, set $\alpha$, $\beta$, $\gamma\in[1,101]$ or $\in[0,1]$,  $m\in \{100,200,300,400,500\}$ to find the optimal parameters of the model. And $k$ varies according to the number of tags. $\alpha$, $\beta$ and $\gamma\in[1,101]$ perform better. In the process of extracting the lower order structure matrix $\textbf{$\mathbb{A}$}$, the dimension $d$ is set to 128. When $m$ is greater than 200, the clustering performance gradually decreases, so the dimension of $m$ is set to 200.

\subsection{Multi-classification experiment}

In this section, the performance of multi-label node classification is evaluated against the metrics MICRO F1 and MACRO F1. In order to reduce the contingency of experimental results, the classification process was repeated 10 times, and the average value was taken as the result. Table \ref{mult_res_1} shows the node classification performance of our model and the baselines of the four datasets for $T = 1$, respectively. All data sets in the same table share the same window size $T$. In these tables, bold numbers indicate the best results.   

\begin{table}[h]
	\caption{Multi-label classification performance evaluation based on Micro/Macro-F1}\label{mult_res_1}
	\vspace{0.5em}\centering
	\begin{tabular}{ccccccccc}
		\toprule[1.5pt]
		\multirow{2}{1cm}{Model}&\multicolumn {2}{c}{BlogCatalog}&\multicolumn {2}{c}{OAG}&\multicolumn {2}{c}{Wikipedia}&\multicolumn {2}{c}{ASN}\\
		& Micro&Macro& Micro&Macro &Micro&Macro &Micro&Macro \\
		\midrule[1pt]
		NetMF&31.54& 14.86& 16.01& 12.10& 49.9& 9.25& 23.86& 4.26\\
		M-NMF& 21.81& 6.53& 18.34& 11.13& 48.13& 7.91& 25.17& 8.66\\
		LINE& 23.74& 13.32& 11.94 &9.54& 41.74& 9.73& 26.06& 7.79\\
		DeepWalk& 29.32& 17.38&12.05& 10.09& 35.08& 9.383& 24.96& 11.87\\
		AROPE& 33.87& 14.51& 19.61& 12.78& \textbf{52.83}& 10.69& 16.23& 10.76\\
		GAE& 27.11& 25.58& 16.67 &11.85& 50.48& \textbf{10.75}& 26.37& 11.02\\	
		\midrule[1pt]
		LNLM&\textbf{34.66}&\textbf{16.19}&\textbf{19.74}&\textbf{13.71}&51.02&9.51&\textbf{26.18}&\textbf{11.69}\\
		\bottomrule[1.5pt]
	\end{tabular}
	\vspace{\baselineskip}
\end{table}

\begin{table}[h]
	\caption{Effects of different T on performance evaluation of multi-label classification based on Micro/Macro-F1}\label{mult_res_T}
	\vspace{0.5em}\centering
	\begin{tabular}{ccccccccccc}
		\toprule[1.5pt]
		\multirow{2}{1.2cm}{$T$}&\multicolumn{2}{c}{Hep-th}&\multicolumn {2}{c}{OAG}&\multicolumn {2}{c}{Wikipedia}&\multicolumn {2}{c}{ASN}\\
		& Micro&Macro& Micro&Macro &Micro&Macro  &Micro&Macro\\
		\midrule[1pt]
		$T$= 2&41.10&25.15&25.47&21.21&53.57&11.98&26.32&13.50\\
		$T$ = 3&41.99&25.98&25.45&21.17&52.75&10.94&26.08&13.58\\
		$T$ = 4&42.01&25.98&25.47&21.22&52.31&10.72&27.24&13.67\\
		$T$ = 5&42.07&26.06&25.88&21.55&51.68&10.47&27.56&13.54\\
		$T$ = 6&42.11&25.89&25.66&21.41&51.25&10.31&27.37&13.72\\
		$T$ = 7&42.13&25.97&25.56&21.38&51.21&10.24&27.26&13.53\\
		$T$ = 8&42.11&25.76&25.60&21.18&50.88&10.07&27.22&13.52\\
		$T$ = 9&42.05&25.81&25.62&21.30&50.64&9.94&27.19&13.51\\
		$T$ = 10&43.15&28.48&25.92&21.59&51.83&10.96&27.13&13.50\\
		\bottomrule[1.5pt]
	\end{tabular}
	\vspace{\baselineskip}
\end{table}
Obviously, according to the evaluation indexes Micro-F1 and Macro-F1, LNLM performs better than other models in the cited network dataset ASN, OAG and Hep-PH, proving the effectiveness of the network embedding model in the analysis of academic networks in this paper. In Wikipedia, the AREOP model shows better performance than the LNLM approach in Micro-F1 and Macro-F1. This phenomenon suggests that a relatively low order is sufficient to characterize the network structure of Wikipedia. The reason is that Wikipedia is a dense word co-occurrence network with a moderate degree of about 85, so if two words appear together in a window of size 2, they will have edges. The results show that the method based on matrix factorization alone does not perform well in the classification task, proving the effectiveness of combining lower-order feature learning node representation on academic networks.

As mentioned earlier, the window size of $T$ determines the order in which structures are captured. In addition, the effect of window size on multi-label classification performance is explored. Here, set the window size $T$ to 1 through 10. The table \ref{mult_res_T}  show the relevant results and trends. Because table \ref{mult_res_1}shows the result of $T$=1, $T$ starts with 2 in table \ref{mult_res_T}. In HEP-TH, OAG, and ASN datasets, the proposed LNLM can significantly better classify performance as the window size $T$ increases, such as Micro-F1 and Macro-F1. However, when  $T$ gradually reaches 4, the performance tends to be stable. In wikis, classification performance gradually degrades when $T$ is greater than 3. This phenomenon indicates that the size of the window will never be as large as possible, and the window should be set dynamically according to the network sparsity to reduce the amount of computation. LNLM model can dynamically adjust the window size according to network sparsity to learn a better node representation.

\subsection{Node clustering experiment}
In this section, the performance of the node cluster is evaluated based on standardized mutual information (NMI) of typical metrics. This paper uses accurate data (including Polbog, Livejournal, and Orkut) to evaluate the clustering performance of real-world datasets. NMI varies between 0 and 1, and the larger the value, the better the cluster performance. In the experiment, the standard K-means algorithm is used to obtain the clustering results of other network embedding methods. Since the initial value significantly influences the clustering results, the clustering is repeated 10 times and its average value is calculated as a result.

Table\ref{t.6} shows node clustering performance concerning NMI. Again, bold numbers indicate the best results in the table. The table results show that LNLM has the best performance on all the network data sets on NMI. In particular, the LNLM approach showed a $24\%$  improvement in NMI compared to the second-best approach on the Pol blog dataset. This is because our method integrates lower-order structure features and local structure features and captures the network's diverse and comprehensive structure features. SDNE and LINE only preserve the proximity between network nodes and cannot effectively maintain the community structure. Deepwalk and Node2vec based on a random walk can capture second-order and even higher similarity. However, they ignore community structure. AROPE can grasp similarities between different nodes. Although more global structure information is caught as the length increases, AROPE still forgets module information. M-NMF introduces modularization items to learn node embedding that preserves community structure. However, for sparse networks and networks with no prominent community structure, the modularization term constraint of NMF makes the representations of nodes similar, so its performance is relatively low. The results show that the method based on matrix factorization only performs poorly in the clustering task. The above results demonstrate the power of fusing lower-order features into embedding while preserving local structures.
\begin{table}[h]
	\caption{Evaluation of node clustering performance based on NMI}\label{t.6}
	\vspace{0.5em}\centering
	\begin{tabular}{cccc}
		\toprule[1.5pt]
		Dataset& Polblog&Orkut& Livejournal\\
		\midrule[1pt]
	    NetMF&0.324&0.557&0.688\\
		M-NMF&0.215&0.310&0.681\\
		DeepWalk&0.475&0.120&0.103\\
		Node2vec&0.453&0.331&0.117\\
		LINE&0.226&0.211&0.565\\
		SDNE&0.077&0.213&0.743\\
		AROPE&0.241&0.306&0.165\\
		GAE&0.369&0.768&0.787\\
		\midrule[1pt]
		\textbf{LNLM}&\textbf{0.718}&\textbf{0.778}&\textbf{0.806}\\
		\bottomrule[1.5pt]
	\end{tabular}
	\vspace{\baselineskip}
\end{table}
\subsection{Link prediction experiment}
In this section's experiment, to predict which node pairs are likely to form a boundary, we hide $10\%$ to $50\%$ of the edges for evaluation as test data while ensuring that the rest of the network is connected. The remaining edges are used to train the node embedding vector, respectively.  A specific area under the curve (AUC) score is used to evaluate the performance of LNLM and other benchmark methods.

First, removing  $10\%$ of the edges on all network datasets to verify the performance of LNLM is shown. As shown in Table \ref{t.7}, the LNLM model achieves $19.6\%$, $7.5\%$ and $9.5\%$ improvements in Pol blog, Orkut and Livejournal, respectively. We note that M-NMF, which preserves the network community structure, is second only to the LNLM model in terms of predictive power in all data sets. 
\begin{table}[!h]
	\caption{Experimental results of link prediction on AUC}\label{t.7}
	\vspace{0.5em}\centering
	\begin{tabular}{ccccc}
		\toprule[1.5pt]
		Dataset& Polblog&Orkut& Livejournal&GRQC\\
		\midrule[1pt]
		NteMF&0.525&0.650&0.806&0.795\\
		M-NMF&0.672&0.835&0.878&0.843\\
		DeepWalk&0.499&0.487&0.469&0.849\\
		Node2vec&0.495&0.516&0.498&0.530\\
		LINE&0.471&0.470&0.515&0.508\\
		SDNE&0.460&0.521&0.529&0.513\\
		AROPE&0.694&0.646&0.775&0.734\\
		GAE&0.859&0.792&0.963&0.937\\
		\midrule[1pt]
		\textbf{LNLM}&\textbf{0.860}&\textbf{0.899}&\textbf{0.972} &\textbf{0.941}\\
		\bottomrule[1.5pt]
	\end{tabular}
	\vspace{\baselineskip}
\end{table}

\begin{figure}[!h]
	\centering
	\includegraphics[width=0.9\textwidth]{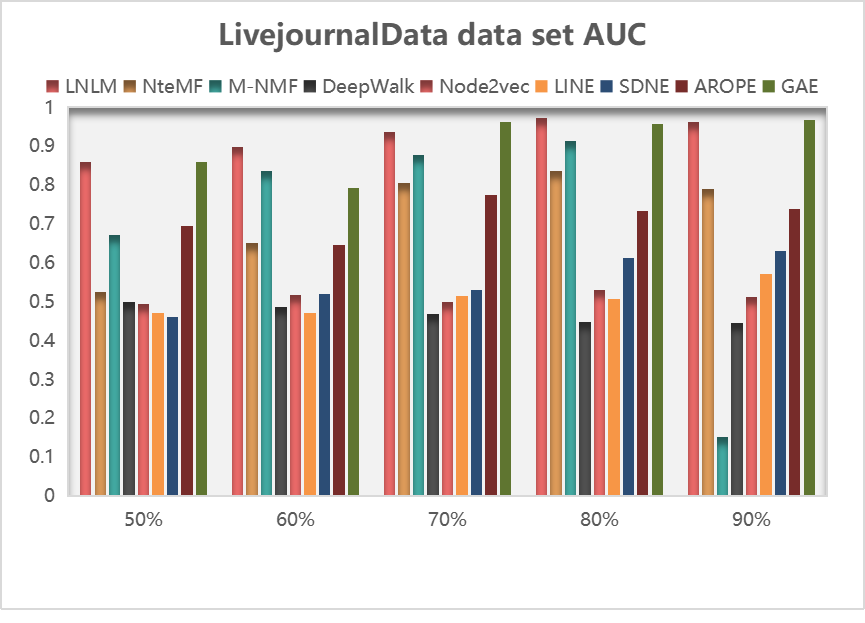}
	\caption{Relation between NMI and M changes}
	\label{fig3}
	\vspace{\baselineskip} %
\end{figure}

Specifically, the results in LiveJournal and Orkut were used as examples to explore the effect of training data ratio. The results in Figure \ref{fig3} show that the LNLM model has a better performance compared to all baselines in both datasets in different parts of the removed edge. Due to the separate network structures, some data sets can achieve the optimal prediction accuracy when $80\%$ of the edges are retained. In comparison, others can achieve the optimal prediction accuracy when $80\%$ of the edges are included. Overall, the results show that the LNLM model can achieve excellent link prediction, indicating the effectiveness of retaining high data sets. Ordered characteristics and local structure information of network embedding.

\subsection{5G academic social network analysis and prediction}
From 2011 to 2014, 4G has just been successfully developed and gradually popularized in China. During this period, the research on the 5G network only stays in the design and application of functions of the 5G network, such as the functional architecture of the mobile network, the future of the architectural design of 5G mobile network, and remote patient monitoring in 5G infrastructure prepared papers to stay in the theoretical writings, the patent application is relatively few. However, after 2018, with the discussion on the construction strategy of the transmission network in the 5G era and the strengthening of the challenges, methods and directions of the 5G network, papers and patents on low-complexity general-purpose filter multiple carriers applicable to the 5G wireless system increased rapidly in this year.

However, with the formal application of 5G network, whether the research hotspot can be further developed, this paper uses the popular ARMA model to predict the number of 5G-related papers and patents.  It can be seen from \ref{fig5}(a) that the data set is a stationary non-white noise sequence, and the ARMA model can model the series. First, we calculate the values of sample autocorrelation coefficient (ACF) and partial correlation coefficient (PACF) of the observation series based on\ref{fig5}(b) and \ref{fig5}(c). Then the ARMA (P, Q) model with appropriate order is selected to fit the properties of sample autocorrelation coefficient and partial autocorrelation coefficient. See the following figure\ref{fig4}:

\begin{figure}[!h]	
	\centering
	\subfigure[Sequence Diagram]{
		\centering
		\includegraphics[width=1\textwidth]{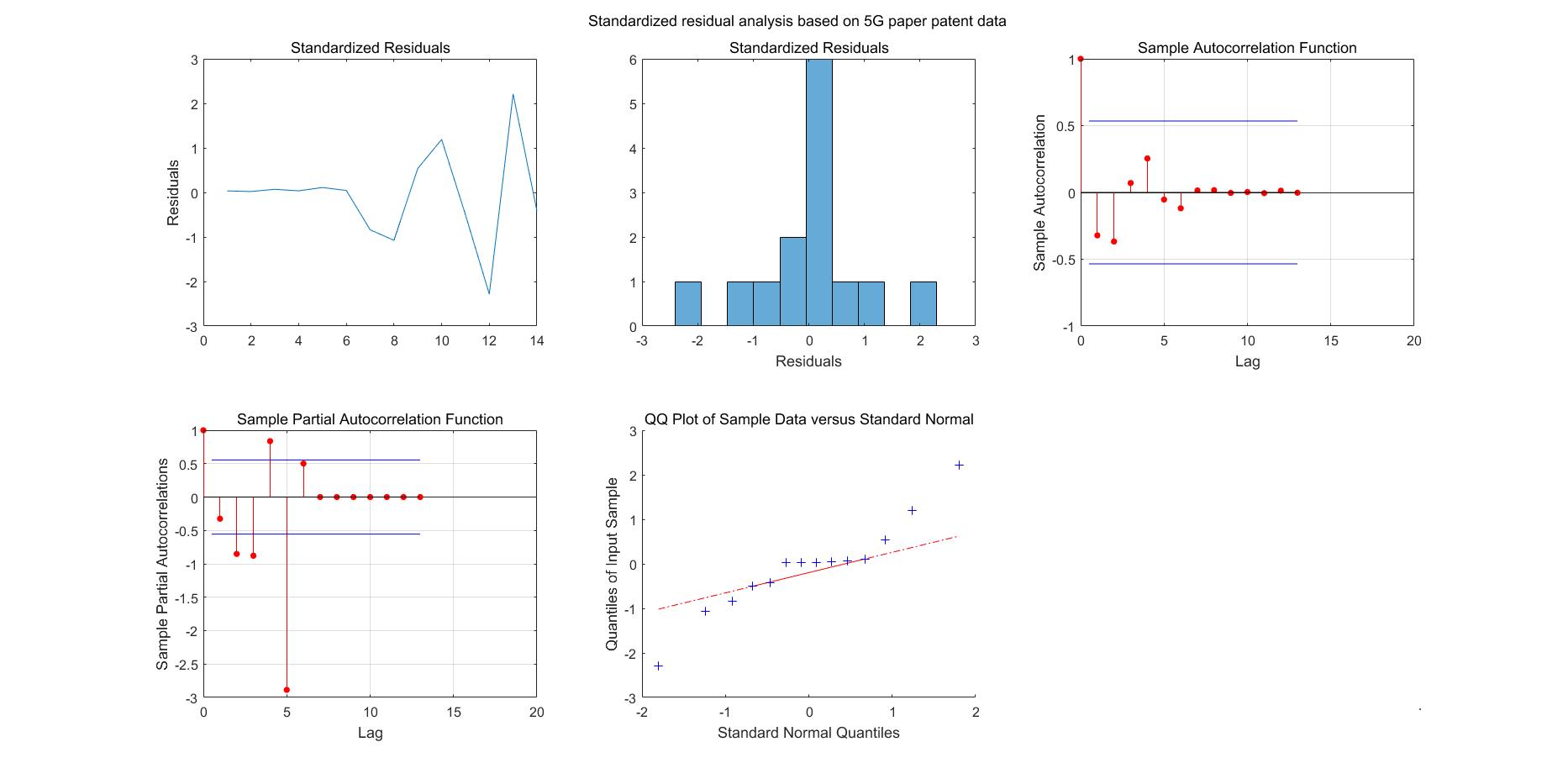}
	}%
    \quad
	\subfigure[ ACF analysis of 5G data]{
		\centering
		\includegraphics[width=2.1in]{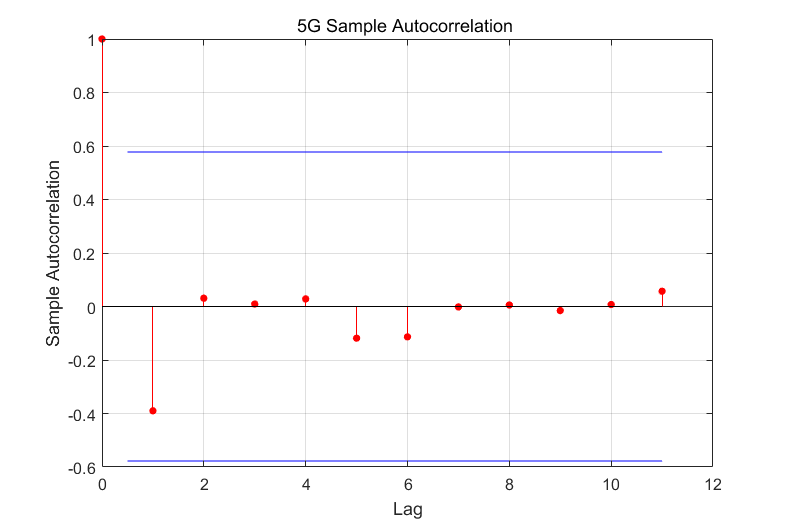}
	}%
	\quad
	\subfigure[ PACF analysis of 5G data]{
		\centering
		\includegraphics[width=2.1in]{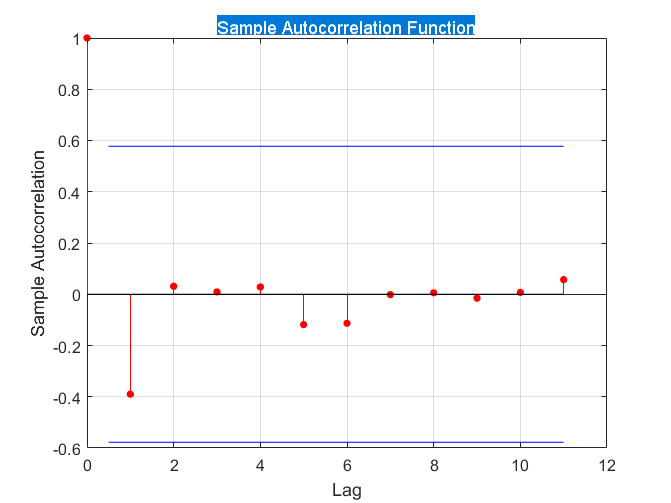}
	}
	\centering
	\caption{ARMA model parameters are determined}
	\label{fig4}
\end{figure}

As shown in \ref{fig4}(b), the PCA autocorrelation diagram shows that all the order autocorrelation coefficients fluctuate within the standard deviation except that the autocorrelation coefficient of order 1 is within the range of 2 standard deviations. The graph has a sinusoidal fluctuation trajectory, which indicates that the attenuation of the autocorrelation coefficient to zero is not a sudden process, but a continuous gradual process. Based on the characteristics of the autocorrelation coefficient, we can judge that the sequence has a short-term correlation and further determine the sequence stability.

\ref{fig4}shows the process of the partial autocorrelation coefficient attenuating to zero. What is unique here is that the partial autocorrelation coefficient of the first order is in the range of 2 standard deviations, and the partial autocorrelation coefficient of the 13th order is also in the range of 5 standard deviations. According to the trailing autocorrelation. Further, we ran the predicted results as shown in the figure \ref{fig5}:
\begin{figure}[!h]
	\centering
	\includegraphics[width=0.9\textwidth]{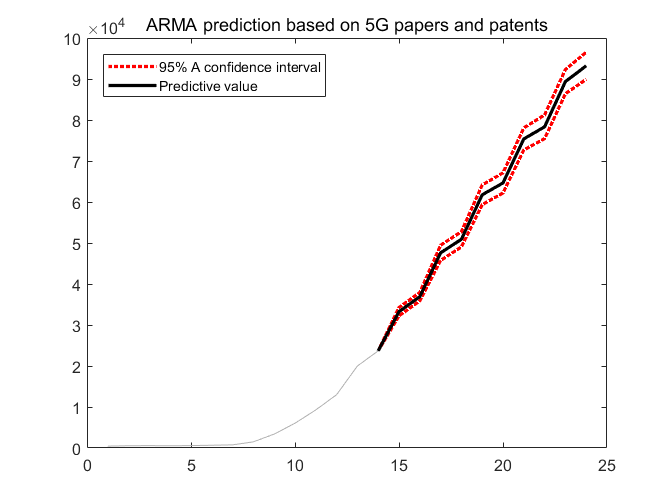}
	\caption{ARMA prediction based on 5G papers and patents}
	\label{fig5}
	\vspace{\baselineskip} %
\end{figure}
As shown in the figure \ref{fig5}, the grey line is the 120 data points used for training, the black line is the prediction of future values, and the red line is the upper and lower limits of the $95\%$ confidence interval. So there's a $95\%$ chance that the actual value of the future will fall within this range.
\section{Industrial applications}
This section uses the LNLM model to verify this 5G academic social network data sets and analysis, through the model after embedding, detection of community and community evolution experiment, 5G academic social network analysis of core were the focus of scientific research team, team leaders and the development direction of different technical fields, and the academic relationship between quoter and development trend.The resulting author collaboration topology is shown the figure \ref{fig6}:
\begin{figure}[!h]
	\centering
	\includegraphics[width=0.9\textwidth]{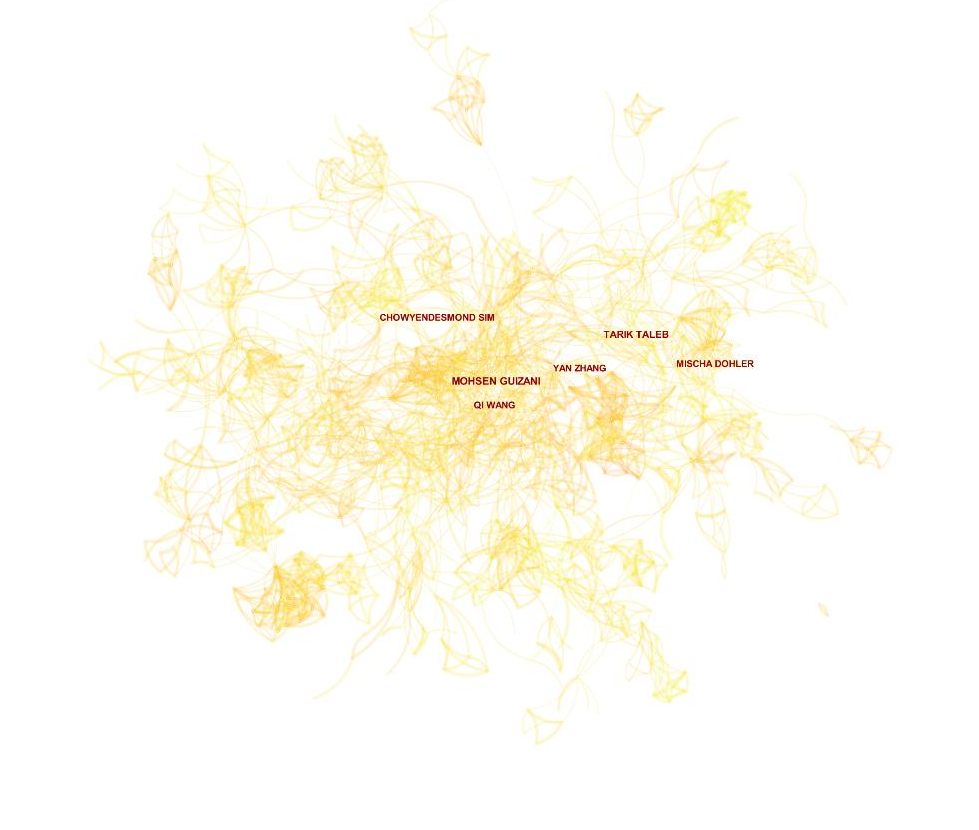}
	\caption{Author collaboration topology}
	\label{fig6}
	\vspace{\baselineskip} %
\end{figure}
Due to many generated nodes, this paper sets labels to display only the top 6 nodes in node degree. The larger the designation of a node is, the greater the degree of the node is. As shown in the figure, Mohsen Guizani's degree value is the largest, so it is the most core node in this network. This is because Mohsen Guizani is a well-known expert in the field of 5G, the chief editor of IEEE fellow, IEEE Network and other international top journals, the University of Idaho, professor of electrical and computer engineering department, research line is wireless communication and mobile computing, computer Network, mobile cloud computing, therefore, Mohsen Guizani has published several high-level papers. The other authors are also experts in the field of 5G, so they occupy a relatively central position in the cooperative network.

\begin{figure}[!h]
	\centering
	\includegraphics[width=0.9\textwidth]{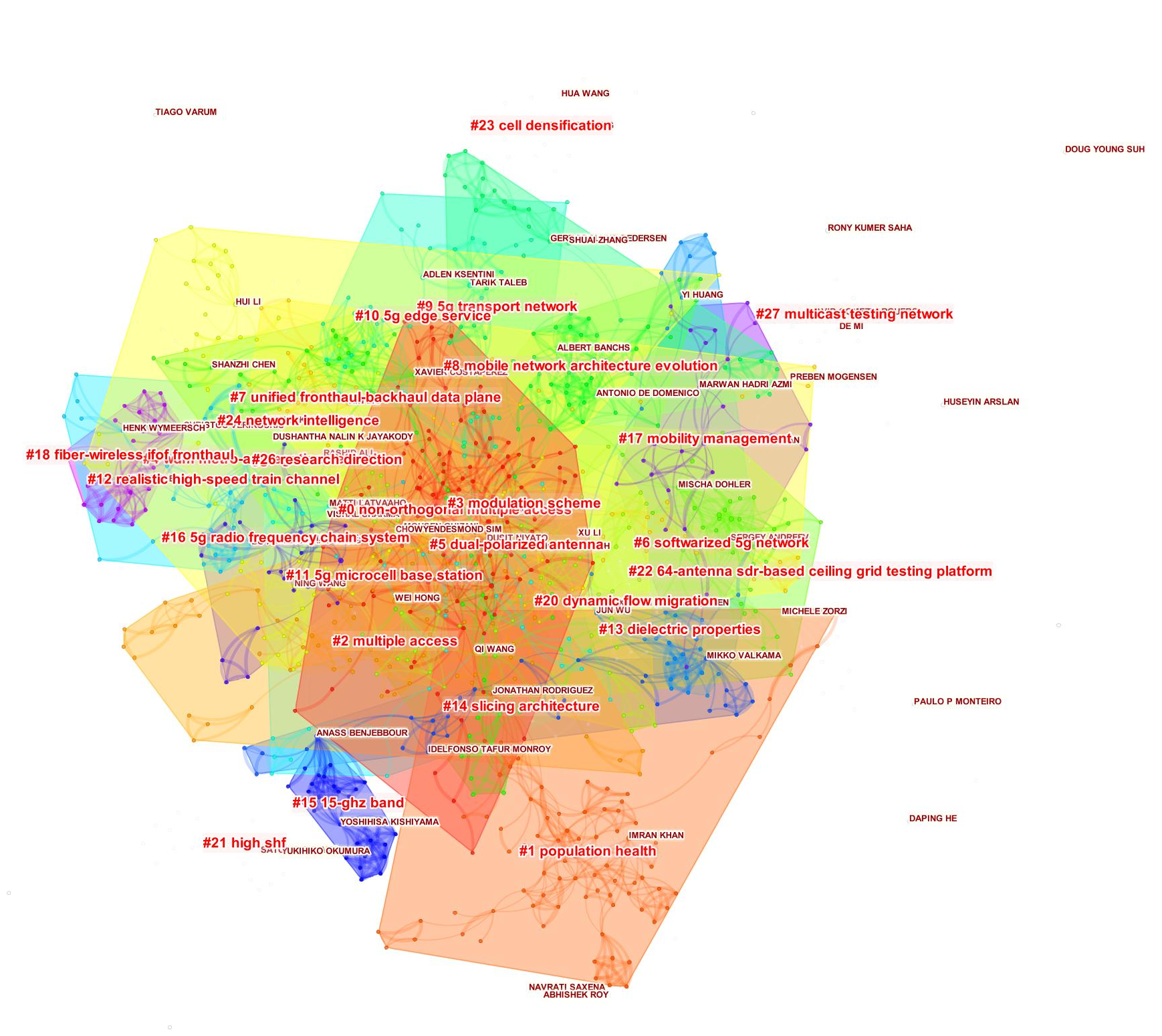}
	\caption{Author Cluster Graph Based on 5G Academic Social Networks}
	\label{fig7}
	\vspace{\baselineskip} %
\end{figure}

According to the figure \ref{fig7}, the author cooperative network can be divided into 26 categories in total. They are Network Intelligence, 5G Microcell Base Station, 5G Transport Network, 5G Edge Service, Mobile Network Architecture Evolution, etc. These categories respectively represent the research direction of the author.

In order to find out the hot research direction of 5G academic social network, this paper selects the top 7 categories of clustering coefficient, Population health, Dual-Polarized antenna, Mobile Network Architecture Evolution, Modulation scheme, Network intelligence, among which network intelligence is the largest category, means that many authors have conducted researches on this research direction, including many well-known professors from Stanford University, Massachusetts Institute of Technology, Tsinghua University and so on. They have collaborated to complete a lot of papers. So that this direction of research personnel, significant influence.
\begin{figure}[!h]
	\centering
	\includegraphics[width=0.9\textwidth]{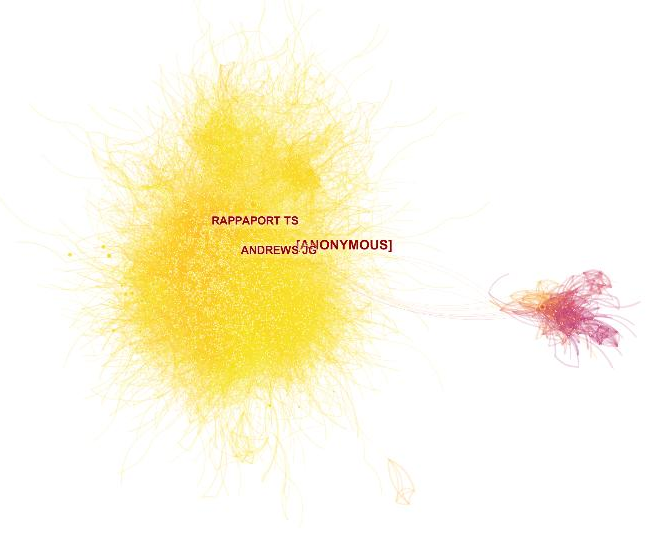}
	\caption{Collaborative topology of cited scholars based on 5G academic social networks}
	\label{fig8}
	\vspace{\baselineskip} %
\end{figure}
In the figure \ref{fig8} shows the cooperation topology of cited scholars, with a total of 26,000 nodes and 125,645 edges. It can be seen from the figure that Anonymous, Rappaport TS, and Andrews JG are highly cited scholars. These three scholars are professors from Harvard University, Stanford University and Oxford University, respectively. Their research direction is 5G intelligence, and they have published many top papers. Therefore, many scholars will appropriately cite their documents in their documents.

\begin{figure}[!h]
	\centering
	\includegraphics[width=0.9\textwidth]{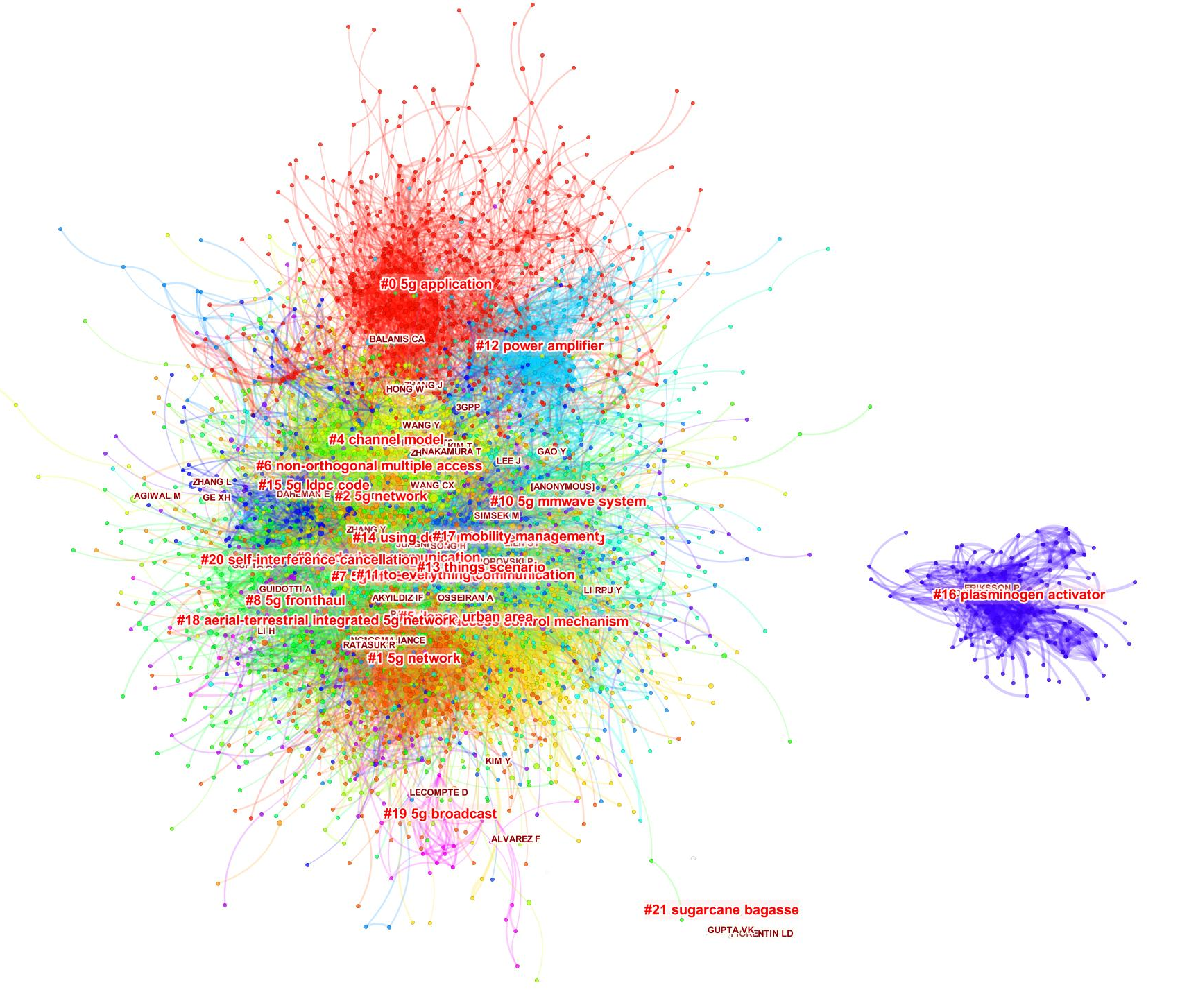}
	\caption{Cited scholar extension based on 5G academic social network}
	\label{fig9}
	\vspace{\baselineskip} %
\end{figure}
In order to better display the cluster diagram of cited 5G scholars, we have built a cooperative network of cited scholars, which has been grouped into 22 categories, including 5G Application, Channel Model, 5G Network, 5G Broadcast, 5G Fronthaul, etc. These all represent the research directions of the cited scholars. We found that they are all related to 5G research, and the core is 5G. According to different structures, they are divided into different directions, namely categories, representing the hot research directions at present.Shown the figure \ref{fig9}:

As shown in the figure \ref{fig9}, 5G application is the category with the most nodes in the clustering, which also indicates that 5G application is the frontier direction of the 5G research field. Many scholars have carried out related researches in this direction. It can be found from the figure that Balans CA, Huang H and Pozah DM have the highest node degree in this category, which means that these three scholars are the leaders in the research direction of 5G application. When we check relevant information, we find that these three scholars are all academicians of the US National Academy of Sciences. I'm a technical engineer with America Mobile.

\section{Conclusions and future work}
With the rapid development of 5G technology, the analysis of 5G academic social networks is of great academic significance and helps guide future scientific development. The diverse forms of collaboration and large scale of data in academic social networks constructed by 5G papers make the management and analysis of academic social networks increasingly challenging. Therefore, this paper builds a low-order feature matrix based on the random walk, and the combined NMF framework allows users to control the weight loss among different structural features. An efficient and scalable network embedding algorithm is proposed. This algorithm can capture the local network structure in the 5G academic social network and effectively integrate the low-order features of nodes into the framework of non-negative matrix factorization to further discover the critical personnel and cooperative communities in the 5G academic social network. The robustness of the proposed algorithm is verified by multi-label classification, clustering and link prediction experiments on four widely used network datasets, three real network datasets and eight mainstream network representation learning models.

It would be great to verify LNLM with a multi-layer graph;  However, we cannot do this in this article due to accessibility issues.  Another problem is that although our model proposes a new approach to building low-order eigenmatrices based on a random walk, it is still challenging to deal with multi-layer sparse data sets.  In future work, we will try to solve this problem through local methods.  


\bibliography{main}

\end{document}